\documentclass{elsart}

\usepackage{natbib}

\usepackage {graphicx}
\usepackage{amssymb}

\usepackage{amsfonts}
\usepackage{latexsym}

\usepackage[latin1]{inputenc}
\usepackage[english]{babel}

\begin{document}

\begin{frontmatter}


\title{Implementing Bayesian predictive procedures:\\
The $K$-prime and $K$-square distributions}

\author[JP]{Jacques Poitevineau\corauthref{cor}}
\corauth[cor]{Corresponding author. Tel: +33-1-5395-4322;
fax: +33-1-4577-1659}
\ead{jacques.poitevineau@upmc.fr}
\author[BL]{Bruno Lecoutre}
\ead{bruno.lecoutre@univ-rouen.fr}

\address[JP]{ERIS and UMR~7190, IJLRA/LAM/LCPE, C.N.R.S.,
Université Pierre et Marie Curie,
11 rue de Lourmel, 75015 Paris, France.}

\address[BL]{ERIS and UPRESA~6085, Laboratoire de Mathématiques
Raphaël Salem, C.N.R.S. et Université de Rouen, Mathématiques,
Site Colbert, 76821 Mont-Saint-Aignan Cedex, France.}

\begin{abstract}
The implementation of Bayesian predictive procedures under standard normal models 
is considered.
Two distributions are of particular interest, the $K$-prime and $K$-square distributions.
They also give exact inferences for simple and multiple correlation coefficients.
Their cumulative distribution functions can be expressed in terms of infinite series of 
multiples of incomplete beta function ratios, thus adequate for recursive calculations.
Efficient algorithms are provided. To deal with special cases where possible underflows 
may prevent recurrence to work properly, a simple solution is proposed which results in 
a procedure which is intermediate between two classes of algorithms. 
Some examples of applications are given.

\end{abstract}

\begin{keyword}
Predictive distribution \sep Bayesian approach 
\sep Incomplete beta function 

\end{keyword}

\end{frontmatter}

\section{Introduction}
Bayesian predictive probabilities give statistical users a particularly
useful device to answer essential questions such as:
``how big should be the experiment to have a reasonable chance of demonstrating a given 
conclusion?'' ``given the current data, what is the chance that 
the final result will be in some sense conclusive, or on the contrary inconclusive?''
Traditional frequentist procedures (e.g., sample size determination via power calculation), 
being conditional to the parameters, are carried out under a subset of paremeter values, 
whereas Bayesian predictive probabilities, which consider all possible parameter values 
(conditionally to the data in hand), give them direct and natural answers.
Some relevant references are 
\citet{Baum}, \citet{Spie}, \citet{John}, \citet{BL01}, \citet{BL08}, \citet{Berr},
\citet{Dmit}, \citet{Grou}.
In particular, from a pilot study, the predictive probabilities on credible limits give 
a useful summary to help in the determination of the sample size of an experiment. 
If the power approach and the predictive approach sometimes result in relatively similar 
sample sizes \citep[for instance, see][]{Inou}, in general, the predictive approach requires 
a larger sample size. This can be considered the price to pay to avoid assumptions about 
parent effect size and variance.

Recently, the Association for Psychological Science has recommended that articles published in 
Psychological Science and their other journals report the ``probability of replicating an effect'', 
denoted $p_{rep}$ \citep{Kila} rather than the traditional $p$-value.
$p_{rep}$ is defined as the predictive probability of finding an effect of the same sign in a 
replication. 

The above procedures are frequently used in the case of comparison of means for which the 
traditional procedures are the $t$ and ANOVA $F$ tests. For sample size determination, 
considering an unknown variance is often seen as an unnecessary sophistication. However, 
this requires that the sample sizes to be determined are relatively high.
The probability of replication $p_{rep}$ -- such as it now appears in Psychological Science -- 
and its extensions frequently involve small sample sizes, but solutions in use assume a 
known variance \citep{BLLP}.One hundred years after Student's famous article \citep{Stud}, 
one can hardly be satisfied with this unnecessary restriction.

The aim of this article is to contribute to implement predictive procedures that relax 
the assumption of a known variance.
These procedures involve the $K$-prime and $K$-square distributions that have been introduced in
\citet{BL84}. They can be characterized as mixtures of the classical noncentral $t$ 
and noncentral $F$ distributions respectively \citep{BL99}.
In particular, the predictive distributions of the $t$ test statistic and the associate limits 
of interval estimates under standard normal models, assuming a conjugate prior, is a 
$K$-prime distribution. The extension to ANOVA $F$ tests involves the $K$-square distribution.
Moerover, the $K$-prime and $K$-square respectively include as particular cases the distributions 
of the sample correlation coefficient and of the sample multiple correlation coefficient, 
alllowing exact inferences about these two coefficients.

This article provide efficient algorithms for the calculation of the cumulative distribution 
functions (cdfs) of these distributions. These cdfs can be expressed in terms of infinite series 
of multiples of incomplete beta function ratios, thus adequate for recursive calculations.
More precisely, both imply the general form
\begin{equation}
\sum_{j=0}^\infty s^jg_jH_j(x),
\label{EQ1}
\end{equation}
with 
\[ s=\pm 1,\quad 0\leq g_j\leq 1\ \forall j,
\quad \sum_{j=0}^\infty g_j=1\]
and where $H_j(x)$ involves only the incomplete beta function.

Dealing with a related problem, the Applied Statistics algorithm
AS 278 developed for the psi-square distribution \citep{psi2} could
be adapted to match the present cdfs. However, AS 278 is a Method~1
recursive algorithm, in the terms of \citet{BK03}: accumulation is
simply done from index $0$ (which maximizes $H_j(x)$) until a 
convergence criterion is met. 
In some cases, especially when the noncentrality parameter of
the distribution is large, it can lead to an exceedingly large number
of iterations, and consequently to unacceptable execution time and
loss of precision. \citet{Frick} proposed an improvement that consists
in starting iterations at an index such that the resulting truncation
error is negligible, but this does not solve the problem.

Yet, the present cdfs are of the general class considered
by \citet{BK03} and, as such, are good candidates for what they called
Method~2 class of algorithms. Essentially, this Method~2 is a
both backward and forward recursive algorithm. As these authors assume 
$\{g_j\}_j$ to be the dominant series in general  (this is 
discussed in section \ref{secPBI}), the starting index
for iterations, say $k$, is chosen so that $g_k$ is a maximum, 
which reduces the above mentioned problems.

Obviously, the best method would be to start iterations at the index
(between $0$ and $k$) which maximizes the product $g_jH_j(x)$ and 
not only one of the terms. However, this is not easy to
determine in general when no analytic solution is available. 
Numerical determination would be time consuming and thus would 
overcome the benefit of an optimal starting point (inasmuch as it should 
be calculated for every $x$). 
We return to this concern in section \ref{secPBI}.

Therefore, we present in the next two sections a Method~2 class of
algorithms applied respectively to the $K$-prime and $K$-square cdfs, 
but of general use as far as the general form (\ref{EQ1}) is concerned. 
In section \ref{secPBI} we compare the two metods and we discuss some remaining 
problems and propose, in some cases, a simple modification which leads to an 
algorithm that is intermediate between Method~1 and Method~2. 
Some examples of applications of these cdfs are given in section 
\ref{secEOA} and section \ref{secCLR} is devoted to some concluding remarks.

\section{$K$-prime distribution} \label{secKP}

Technical characterizations of the $K$-prime distribution can be found
in \citet{BL99}. This distribution is written $K'_{q,r}(a)$ where
$q, r$ are degrees of freedom parameters and $a$ is a noncentrality
parameter.

Particular cases of the $K$-prime distributions are:\\
$a=0$ : \ $K'_{q,r}(0) \equiv t_r$ (usual $t$ distribution),\\
$q=\infty$ : \ $K'_{\infty,r}(a) \equiv t'_r(a)$ (noncentral $t$
distribution),\\
$r=\infty$ : \ $K'_{q,\infty}(a) \equiv \Lambda'_q(a)$
(lambda-prime distribution),\\
$q=\infty,r=\infty$ : \ $K'_{\infty,\infty}(a) \equiv N(a,1)$
(normal distribution).\\

This cdf has the following properties:
\[\Pr(K'_{q,r}(a)<x)= \Pr(K'_{r,q}(x)>a), \]
\[\Pr(K'_{q,r}(-a)<-x)= \Pr(K'_{q,r}(x)>a),\]
\[\Pr(K'_{q,r}(a)<0)= \Pr(\Lambda'_q(a)<0)=\Pr(t_q>a).\]

Several cases are to be distinguished for the cdf:\\
\\
If $a>0$ and $x<0$
\begin{eqnarray*}
\Pr(K'_{q,r}(a)<x)&=& \Pr(K'_{q,r}(a)<0)-\Pr(x<K'_{q,r}(a)<0)\\
   &=& \Pr(t_q>a)-\sum_{j=0}^{\infty} (-1)^jg_jH_{j}(x),\
\end{eqnarray*}
where
\[
g_j=\frac{1}{2}\frac{\Gamma(\frac{q+j}{2})}{\Gamma(1+\frac{j}{2})
\Gamma(\frac{q}{2})}\left(\frac{q}{q+a^2}\right)^{\frac{q}{2}}
\left(\frac{a^2}{q+a^2}\right)^{\frac{j}{2}},
\]
\[
H_j(x)=I_{x^2/(r+x^2)}\left(\frac{j+1}{2}, \frac{r}{2}\right),\
\]
and $I_z$ is the incomplete beta function
\[
I_z(a,b)=\frac{\Gamma(a+b)}{\Gamma(a)\Gamma(b)}
\int_{0}^{z} t^{a-1}(1-t)^{b-1}dt.
\]
If $a>0$ and $x>0$
\begin{eqnarray*}	
\Pr(K'_{q,r}(a)<x)&=& \Pr(K'_{q,r}(a)<0)+\Pr(0<K'_{q,r}(a)<x)\\
   &=& \Pr(t_q>a)+\sum_{j=0}^{\infty} g_jH_{j}(x).
\end{eqnarray*}
If $a<0$, we reduce to the above cases using
\[\Pr(K'_{q,r}(a)<x) = 1-\Pr(K'_{q,r}(-a)<-x).\]
If $a=0$, we simply have
\[\Pr(K'_{q,r}(0)<x) = \Pr(t_r<x).\]

Hence, the cdf of the $K$-prime involves the calculation of the cdf of
the usual Student's $t$ distribution and a series of the general form
(\ref{EQ1}). The case where $a$ and $x$ are of a different sign is an
unfavorable one, since the series is then alternate. Therefore, in the
algorithm, the even and odd terms of the series should be  
accumulated separately in order to minimize the number of subtractions.

The forward and backward recurrence relations for the cdf are straightforward. 
For the $H_j$'s (the incomplete beta function) we have
\begin{eqnarray*}
H_{j+2}&=&H_j-\frac{\Gamma(\frac{j+r+1}{2})}
{\Gamma(\frac{j+3}{2})\Gamma(\frac{r}{2})}
\ \left( \frac{x^2}{r+x^2}\right)^{\frac{j+1}{2}}
\left(\frac{r}{r+x^2}\right)^\frac{r}{2},\\
H_{j-2}&=&H_j+\frac{\Gamma(\frac{j+r-1}{2})}
{\Gamma(\frac{j+1}{2})\Gamma(\frac{r}{2})}
\ \left( \frac{x^2}{r+x^2}\right)^{\frac{j-1}{2}}
\left(\frac{r}{r+x^2}\right)^\frac{r}{2}
\end{eqnarray*}
and for the $g_j$ coefficients
\begin{eqnarray*}
g_{j+2}&=&\frac{q+j}{j+2}\ \frac{a^2}{q+a^2}\ g_j, \\
g_{j-2}&=&\frac{j}{q+j-2}\ \frac{q+a^2}{a^2}\ g_j.
\end{eqnarray*}

From the forward recurrence relation, it is straightforward that imposing 
$g_{j+2}<g_j$ leads to $j > a^2(q-2)/q-2$. Thus, the starting point for iterations, 
say $k$, is taken as the mode of the $g_j$'s, i.e. $k=[a^2(q-2)/q]-1$, 
where $[.]$ denotes the integer part. 

Given the parameters, $H_j(x)$ is a decreasing function of $j$.
Thus, when stopping the calculations at step $j$, the truncation
error ($E_t$) is bounded by: \\
\\
while $j < k$
\begin{eqnarray}
E_t &\leq& H_0(x)\sum_{i=0}^{k-j-1}g_i +
			H_k(x)\sum_{i=k+j+1}^\infty g_i \nonumber \\
    &\leq& H_0(x)\sum_{i=0}^{k-j-1}g_i +
    	H_0(x)\sum_{i=k+j+1}^\infty g_i \nonumber \\
    &\leq& H_0(x)\left[ 1-\sum_{i=k-j}^{k+j}g_i\right] \label{truncH0}
\end{eqnarray}
and when $j \geq k$
\[
E_t \leq H_{k+j}(x)\left[ 1-\sum_{i=0}^{k+j}g_i\right].
\]

\citet{BK03} used $E_t \leq1-\sum_{i=k-j}^{k+j}g_i$ instead of (\ref{truncH0}),  
so that the calculation of $H_0(x)$ was avoided. We think that the relaxation of 
the stopping rule compensates for the increased execution time due to one call to 
the incomplete beta function.

\emph{Stopping rule}: Stop when $E_t$ becomes lower than a
predetermined absolute error bound.

\section{$K$-square distribution} \label{secK2}

Technical characterizations of the $K$-square distribution can be found
in \citet{BL99}. This distribution is written $K^2_{p,q,r}(a^2)$
where $p, q, r$ are degrees of freedom parameters and $a^2$ is a
noncentrality parameter.

Particular cases of the $K$-square distribution are:\\
$a=0$ : \ $K^2_{p,q,r}(0) \equiv F_{p,r}$ (usual $F$ distribution),\\
$q=\infty$ : \ $K^2_{p,\infty,r}(a^2) \equiv F'_{p,r}(a^2)$
(noncentral $F$ distribution),\\
$r=\infty$ : \ $K^2_{p,q,\infty}(a^2) \equiv \Lambda_{p,q}^2(a^2)$
(lambda-square or alternate chi-square distribution),\\
$q=\infty, r=\infty$ : \ $K^2_{p,\infty,\infty}(a^2)
\equiv (1/p)\chi_p^2(a^2)$ (noncentral chi-square distribution).\\

For the cdf, $s=1$ in (\ref{EQ1}) and we simply have
\[ \Pr(K_{p,q,r}^2(a^2)<x)=\sum_{j=0}^{\infty} g_jH_j(x), \]
with
\[
g_j=\frac{\Gamma(\frac{q}{2}+j)}{\Gamma(j+1)\Gamma(\frac{q}{2})}
\left(\frac{q}{q+a^2}\right)^{\frac{q}{2}}
\left(\frac{a^2}{q+a^2}\right)^j
\]
and
\[
H_j(x)=I_{px/(r+px)}\left(\frac{p}{2}+j, \frac{r}{2}\right),\ x>0,
\]

The recurrence relations for the incomplete beta function now write
\begin{eqnarray*}
H_{j+1}&=&H_j-\frac{\Gamma(p/2+r/2+j)}{\Gamma(p/2+j+1)\Gamma(r/2)}
\ \left(\frac{px}{r+px}\right)^{p/2+j}
\left(\frac{r}{r+px}\right)^{r/2},\\
H_{j-1}&=&H_j+\frac{\Gamma(p/2+r/2+j-1)}{\Gamma(p/2+j)\Gamma(r/2)}
\ \left(\frac{px}{r+px}\right)^{p/2+j-1}
\left(\frac{r}{r+px}\right)^{r/2}
\end{eqnarray*}
and for the $g_j$ coefficients
\begin{eqnarray*}
g_{j+1}&=&\frac{q/2+j}{j+1}\ \frac{a^2}{q+a^2}\quad g_j,\\
g_{j-1}&=&\frac{j}{q/2+j-1}\ \frac{q+a^2}{a^2}\quad g_j.
\end{eqnarray*}

The coefficients $g_j$ are the probabilities of obtaining the value
$j$ for a variate following a negative binomial distribution with
parameters $q/(q+a^2)$ and $q/2$. The mode is $[a^2(q-2)/(2q)]$ 
(where $[.]$ denotes the integer part),
hence the starting index for iterations. The stopping rule is the same 
as in the case of the $K$-prime.

\section{Limitations and possible improvements} \label{secPBI}

Drawbacks of Method~1 algorithms (possible underflows and an exceeding number 
of iterations) led to the development of Method~2 algorithms. 
In Method~1, the iterations start at index $j=0$ which maximizes $H_j(x)$, 
while in Method~2 they start at index $j=k$ which maximizes $g_j$. 

In Tables 1 and 2, we compare the number of iterations for 
these two methods, as applied respectively to the $K$-prime and $K$-square 
cdfs for various situations and with a precision set to $10^{-4}$. 
The ten first examples in Table~2 correspond to those in Table~1 of \citet{BK03} 
for the distribution of the square of the sample multiple correlation 
coefficient. More precisely, the correspondence is as follows:
the sampling distribution of the multiple correlation $R^2$, involving a
sample of $n$ independent observations from a $m$-variate normal
population with square multiple correlation coefficient $\rho^2$,
is such that
\[
\frac{n-m}{r-1}\ \frac{R^2}{1-R^2}\;|\;\rho^2
\sim K^2_{m-1,n-1,n-m}\left((m-1)\frac{\rho^2}{1-\rho^2}\right).
\]

One last example has been added, corresponding to $r^2=0.33$, $\rho^2=0.50$, $m=5$, $n=100$. 

Of course, as soon as both methods attain at least $2k$ iterations, they return identical results 
as the same terms are summed up (for instance, this is the case in the fifth example of Table~1). 
As can be seen, relatively to Method~1, Method~2 can indeed reduce the number of 
iterations by a great amount: more than 60\% in most of Table~2 examples. 
When the precision criterion is turned to $10^{-12}$ \citep[as in][]{BK03}, 
the gain diminishes, naturally, but is stil about 40\%.
However, it is also obvious that Method~2 is not systematically better. This can be seen 
in the last example of Table~2, and is especially clear in the case of the $K$-prime distribution 
(Table~1) where the number of iterations can be increased by more than 1000\%. 
It's not surprising that Method~1 performs better when the noncentrality parameter is small, 
but it also happens when this parameter is higher, as in the case of the second and third 
examples of Table~1.

More generally, whenever$H_k(x)$ tends to zero quickly with respect to $k$, 
Method~1 algorithms perform better than Method~2 algorithms, because only 
the first terms of the series (\ref{EQ1}) contribute significantly to 
the sum. And when $H_k(x)$ is still close to $H_0(x)$, Method~2 is likely 
to be quasi optimum.

\begin{table}
\caption{Comparison between Methods 1 and 2 
for the $K$-prime cdf algorithm. M$i$ is number of iterations for Method~$i$. 
Gain is the gain, in percentage, of Method~1 over Method~2, a negative number 
indicates Method~1 performs better.}
\begin{tabular}{l l l l l l l l}
 \hline
 $x$&$q$&$r$&$a$&$\Pr(K'_{p,q,r}(a)<x)$&M1&M2&gain\\
 \hline
 1&   5&20&10& 0.0007&   9& 119&-1222\% \\
11&   5&20&50& 0.0017& 332&2999& -803\% \\
40&  50&50&50& 0.0612&2892&4799&  -60\% \\ 
40&  50& 5&50& 0.4277&4387&4799&   -9\% \\ 
50&  50&20&30& 0.5242&1844&1844&    0\% \\
40& 100& 5&50& 0.1783&3644&3224&   12\% \\
45& 100&10&40& 0.6377&2499&2084&   17\% \\
65&1000&15&50& 0.8820&3007&1052&   65\% \\
 \hline
\end{tabular}
\end{table}

\begin{table}
\caption{Comparison between methods 1 and 2 
for the $K$-square cdf algorithm. M$i$ is number of iterations for Method~$i$. 
Gain is the gain, in percentage, of Method~1 over Method~2, a negative number 
indicates Method~1 performs better.}
\begin{tabular}{l l l l l l l l l}
 \hline
 $x$&$p$&$q$&$r$&$a^2$&$\Pr(K_{p,q,r}^2(a^2)<x)$&M1&M2&gain\\
 \hline
     36& 2&  20&  18&46.667& 0.7771&   57&  57&  0\% \\
0.19444& 4&  11&   7&4.7143& 0.0126&    3&   3&  0\% \\
    288& 3&  99&  96&   891& 0.4382&  618& 598&  3\% \\
    972&11&1199&1188& 10791& 0.4339& 5953&1844& 69\% \\
  795.2& 5& 999& 994&  3996& 0.4661& 2246& 796& 65\% \\
  475.2& 5& 599& 594&  2396& 0.4562& 1390& 624& 65\% \\
  715.2& 5& 899& 894&  3596& 0.4643& 2033& 756& 63\% \\
202.909&11&1499&1488&2248.5& 0.4297& 1252& 420& 66\% \\
216.545&11&1599&1588&2398.5& 0.4319& 1331& 433& 67\% \\
223.364&11&1649&1638&2473.5& 0.4330& 1371& 439& 68\% \\
11.6978& 4&  99&  95&    99& 0.0063&   47&  90&-91\% \\
 \hline
\end{tabular}
\end{table}

Furthermore, with Method~2, it can happen that the initial recurrence increment 
for the $H_j$'s is too small with respect to the machine limit so that a zero is 
returned and recurrence is impossible: e.g., for the $K$-square cdf, this increment 
term is lower than $10^{-307}$ when $p=10, q=20, r=30, a^2=500$ and $x=0.1$. 
So, both methods are subject to underflows, whether through the $g_j$'s (Method~1) or 
whether through the $H_j(x)'s$ (Method~2). 

A tempting solution, when $H_k(x)$ is too small, would be to choose 
a modified index, say $k'$, such that $H_{k'}(x)$ reaches a predetermined value 
(i.e. one markedly above the machine limit); unfortunately, such an inversion of the beta 
cdf involves an iterative procedure and so is to be discarded on grounds of speed 
efficiency. An alternative solution is to lower $k$ by some amount. This amount will 
depend, among others, on $x$. Given the parameters and $j$, $H_j(x)$ is an increasing 
function of the argument of the incomplete beta function, say $z$, that is itself 
a function of $x$. The lower $H_j(x)$, the more $k$ has to be lowered. Thus, for 
sake of simplicity and as a first attempt, we propose to use the identity function 
on $z$ so that $k$ is simply lowered by multiplying it by the argument of the incomplete 
beta function ($px/(px+r)$ for the $K$-square and $x^2/(x^2+r)$ for the $K$-prime).

This modification avoids underflows in the preceding example. Furthermore, it sometimes 
permits to reduce the number of iterations. Thus, it could be introduced as soon as  
$H_k(x)$ is below some arbitrary threshold (e.g., when $H_k(x)/H_0(x)<0.01$) and not 
only when a true underflow occurs.
For instance, for the distribution $K^2_{10,80,200}(500)$, when $x$ takes the values 
35, 30, 20, and 10, the number of iterations is always 390 (for a precision of $10^{-4}$), 
while when turning to the modified starting index, it drops respectively to 309, 291, 
243 and 163. In the first example of Table~1, the modification leads to 
9 iterations (instead of 119 with the unmodified version), just as Method~1. 
Obviously, it is not relevant when $H_j(x)$ diminishes rather slowly with $j$, which is 
the case for the Table~2 examples, except the last one. In that last example, the 
modification leads again to the same number of iterations as Method~1 (47).
Another example of reduction of iterations, concerning the $K$-prime distribution, 
is given in the next section.

Therefore, we could finally suggest the following tactic:
\begin{enumerate}
	\item Calculate $g_0H_0(x)$ and $g_kH_k(x)$ and choose as the starting index 
($0$ or $k$) the one which leads to the maximum.
	\item If $0$ is chosen and recurrence is impossible, try $k$.
	\item If $k$ is chosen and recurrence is impossible (or if $H_k(x)$ is very 
	small compared to $H_0(x)$), multiply it by the argument of the incomplete 
	beta function (this can be repeated).
\end{enumerate}

\section{Examples of applications} \label{secEOA}

\subsection{Predictive probabilities} 

Suppose a simple two-sample experiment was designed to compare a new drug with a placebo. 
For this purpose, the investigators used a two-sample $t$ test with equal numbers of 
subjects $n_1=10$ in each group, in order to test 
H$_0: \delta=0$ against the alternative H$_1: \delta>0$. 
Let us denote by $m_1$ the sample mean difference in the current data and by $s_1$ the pooled
estimate of the common standard deviation $\sigma$. 
The observed $t$ statistic was $T_1=1.10$, hence $p=0.143$ (one-tailed).

Let us consider a conjugate prior for ($\mu,\sigma^2$), characterized by
\[
\mu|\sigma^2 \sim N(m_0,\frac{2}{n_0}\sigma^2) \hbox{ and } 
	\sigma^2 \sim s_0^2\bigg(\frac{\chi_{q_0}^2}{q_0}\bigg)^{-1}.
\]

\noindent \citet{BL99} demonstrated that the predictive distribution of the $t$ test 
statistic for $n$ future observations is a $K$-prime distribution

\[t \sim \sqrt{1+\frac{n}{n_0}}\, K'_{q_0,2n-2}\left(\frac{T_0}{\sqrt{1+\frac{n_0}{n}}}\right)
 ~~~\hbox{where } T_0 = \frac{m_0}{s_0}\sqrt{n_0/2}.\]

As a particular case, here the prior is the posterior distribution from the available data, 
starting with the usual noninformative prior $p(\mu,\sigma^2) \propto 1/\sigma^2$,
hence $m_0$ = $m_1$, $s_0=s_1$, $n_0=n_1$ and $q_0=2n_1-2$. 
We get for a replication ($n=n_1=10$) the predictive distribution

\[t \sim \sqrt{2}\,K'_{18,18}\bigg(\frac{T_1}{\sqrt{2}}\bigg),\]

\noindent that only depends on the observed $t$ test statistic $T_1=1.10$ and its 
associated degrees of freedom.

We can compute the probability of finding a positive mean in a replication 
(Killeen's $p_{rep}$) as

\[Pr\bigg(K'_{18,18}\Big(\frac{1.10}{\sqrt{2}}\Big) > 0\bigg) = 0.777.\]

We can also compute the predictive probability of a significant replication. For instance 
we find the probability 0.334 that the one-tailed $p$ value will be less than 0.05 
(i.e. $t>1.734$): 

\[Pr\bigg(\sqrt{2}\,K'_{18,18}\Big(\frac{1.10}{\sqrt{2}}\Big) > 1.734\bigg)
	= Pr\bigg(K'_{18,18}\Big(\frac{1.10}{\sqrt{2}}\Big) > 1.226\bigg) = 0.334.\]

The investigators generally largely underestimate this probability: see \citet{MPLR}, \citet{MPL0}.
Note that there is also a non negligible probability of finding a significant effect 
in the negative direction: $Pr(\sqrt{2}\,K'_{18,18}(1.10)/\sqrt{2}) < -1.734) = 0.027$. 

An example of application to sample size determination in clinical trials from a pilot study 
is given in \citet{Grou}.

The predictive probabilities for $F$ ratios and usual standardized effect size measures in 
ANOVA designs can be computed from the $K$-square distribution.
Let us consider for instance the data of a one-way design with $g$ groups of equal sample sizes $n_0$. 
Let $F_0$, the observed ANOVA $F$ ratio for the overall comparison of the means.
Assuming before the experiment the usual non informative prior, the posterior predictive 
distribution for the $F$ ratio in a future experiment with equal sample sizes $n$ is a 
$K$-square distribution \citep{BL99}:

\[ F \sim \frac{1+\frac{n}{n_0}}{g-1}\ 
K^2_{g-1,gn_0-g,gn-g} \left(\frac{g-1}{1+\frac{n_0}{n}}F_0\right).\]

\subsection{Distributions of correlation coefficients} 

Other applications of the $K$-prime and $K$-square distributions are exact inferences
for correlation coefficients.
The sampling distribution of the correlation coefficient $r$, involving a
sample of $n$ independent observations from a bivariate normal
population with population coefficient $\rho$, is such that

\[\sqrt{n-2}\,\frac{r}{\sqrt{1-r^2}}\;|\;\rho
\sim K'_{n-1,n-2}\left(\sqrt{n-1}\,\frac{\rho}{\sqrt{1-\rho^2}}\right),\]

\noindent so that exact tests and confidence limits for $\rho$ can be computed from the 
$K$-prime cdf. For instance, when $n=250$ and assuming $\rho=0.80$, the probability to observe  
a sample $r$ lower than $0.75$ is 0.0227. If this is calculated using the standard 
Method~2 algorithm with a precision of $10^{-12}$, 860 iterations are required, whereas 
only 595 are needed with the modification proposed in section \ref{secPBI} (if the 
precision is set to $10^{-6}$, the numbers of iterations become respectively 546 and 502).

Moreover, in the Bayesian framework, assuming a uniform prior for $\rho$, the posterior  
distribution is also a $K$-prime distribution. 

The sampling distribution of the multiple correlation $R^2$ has been presented 
in section \ref{secPBI}.

\section{Concluding remarks} \label{secCLR}

We presented an algorithm for two Bayesian predictive distributions
involved in the designing of experiments and in the computation of 
``the probability of replication'' under usual normal models.
Furthermore, we used these distributions to compare two available methods 
for computing cdfs that are expressed as discrete mixtures of continuous 
distributions (the incomplete beta function in our case). 
If in many cases the two methos are likely to perform equally well, it appeared that 
none of them is systematically better, depending, among others, upon the particular functions 
involved in the cdfs, and that they both suffer a comparable problem: 
due to underflows, the starting index of iterations can be such that recurrence 
is impossible. 
Method~2 was proposed to avoid Method~1 underflows, and here we proposed to manage 
Method~2 underflows by lowering the starting index by a quantity which is the 
argument of the incomplete beta function. This is a tentative solution that can be 
viewed as a crude approach to the problem of finding the optimum starting index.

\ack

The authors wish to thank an anonymous referee whose thoughtful comments 
permitted to greatly improve the manuscript.

\end{document}